\documentclass[graybox]{svmult}

\usepackage{mathptmx}       
\usepackage{helvet}         
\usepackage{courier}        
\usepackage{type1cm}        
\usepackage{makeidx}         
\usepackage{graphicx}        
\usepackage{multicol}        
\usepackage[bottom]{footmisc}

\makeindex             

\begin{document}

\title*{Search for Turbulent Gas through Interstellar  Scintillation}

\author{Moniez M., Ansari R., Habibi F., Rahvar S.}

\institute{Moniez \at Laboratoire de l'Acc\'{e}l\'{e}rateur Lin\'{e}aire,
{\sc IN2P3-CNRS}, Universit\'e de Paris-Sud, B.P. 34, 91898 Orsay
Cedex, France, \email{moniez@lal.in2p3.fr}
}
\maketitle

\abstract{
Stars twinkle because their light propagates through the atmosphere.
The same phenomenon is expected when the light of remote stars
crosses a Galactic -- disk or halo -- refractive medium
such as a molecular cloud.
We present the promising results of a test performed with the ESO-NTT
and the perspectives.
\keywords{Galaxy:structure, dark matter, ISM}
}
\section{What is interstellar scintillation?}
\begin{figure}[h]
\begin{center}
\includegraphics[width=11.cm]{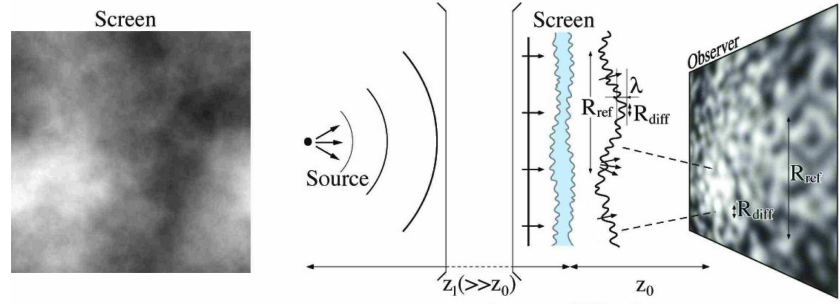}
\end{center}
\caption[] 
{\it
Left: a 2D stochastic phase screen (grey scale), from
a simulation of gas affected by Kolmogorov-type turbulence.

Right: the illumination pattern from a point source (left) after crossing
such a phase screen.
The distorted wavefront produces structures at scales $\sim R_{diff}(\lambda)$
and $R_{ref}(\lambda)$ on the observer's plane.
}
\label{front}
\end{figure}

Refraction through
an inhomogeneous transparent cloud (hereafter called screen)
distorts the wave-front of incident electromagnetic waves
(Fig.\ref{front}) \cite{Moniez};
For a {\it point-like} source,
the intensity in the observer's plane is affected by
interferences which, in the case of stochastic inhomogeneities,
takes on the speckle aspect.
At least 2 distance scales characterise this speckle:
\begin{itemize}
\item
The diffusion radius $R_{diff}(\lambda)$ of the screen,
defined as the transverse separation
for which the root mean square of the phase difference at wavelength
$\lambda$ is 1 radian.
\item
The refraction radius
\begin{equation}
R_{ref}(\lambda)=\frac{\lambda z_0}{R_{diff}(\lambda)} \sim
30860\, km \left[\frac{\lambda}{1\, \mu m}\right]
\left[\frac{z_0}{1\, kpc}\right]\left[\frac{R_{diff}(\lambda)}{1000\, km}\right]^{-1}
\end{equation}
where $z_0$ is the distance to the screen.
This is the size, in the observer's plane, of the diffraction spot from a patch
of $R_{diff}(\lambda)$ in the screen's plane.
\end{itemize}
After crossing a fractal cloud described by the Kolmogorov turbulence
law (Fig. \ref{front}, left), the light from a {\it monochromatic
point} source produces an illumination
pattern on Earth made of speckles of size $R_{diff}(\lambda)$ within
larger structures of size $R_{ref}(\lambda)$ (Fig. \ref{front}, right).

\begin{figure}[h]
\centering
\parbox{8cm}{
\includegraphics[width=8cm]{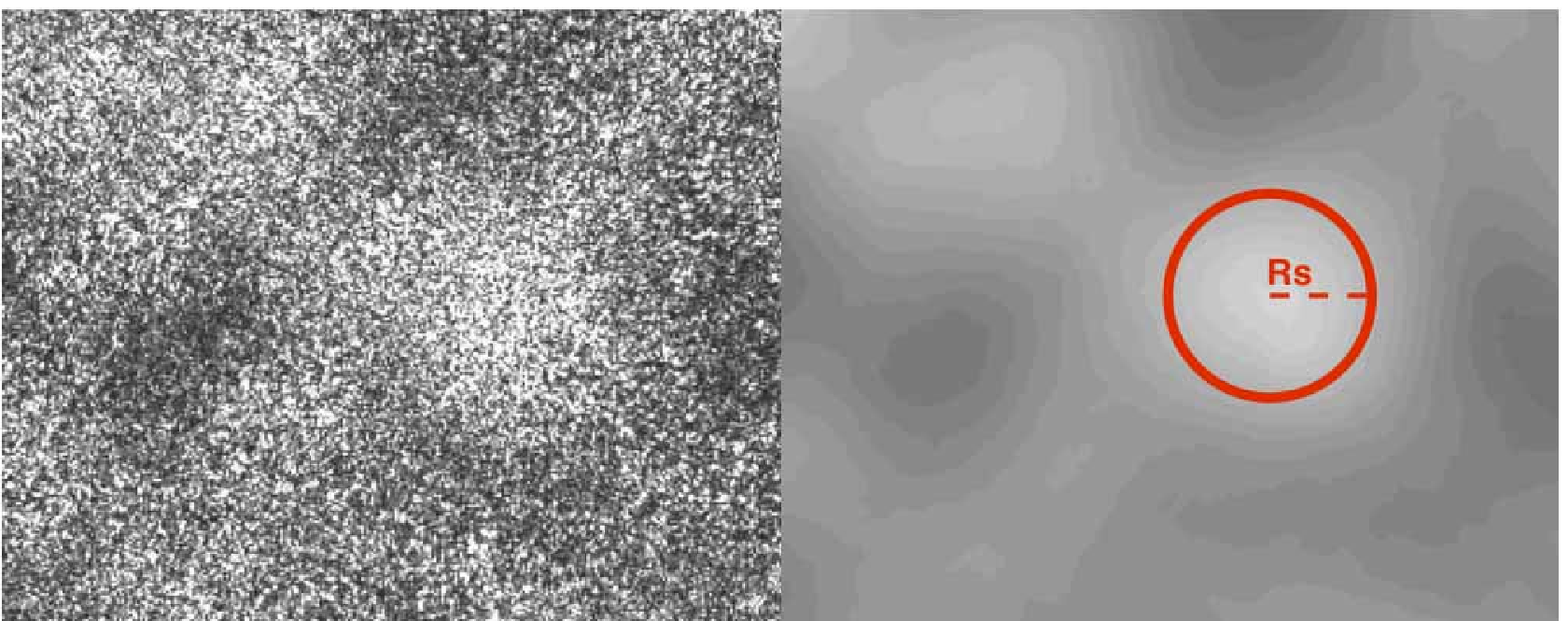}
\includegraphics[width=8cm]{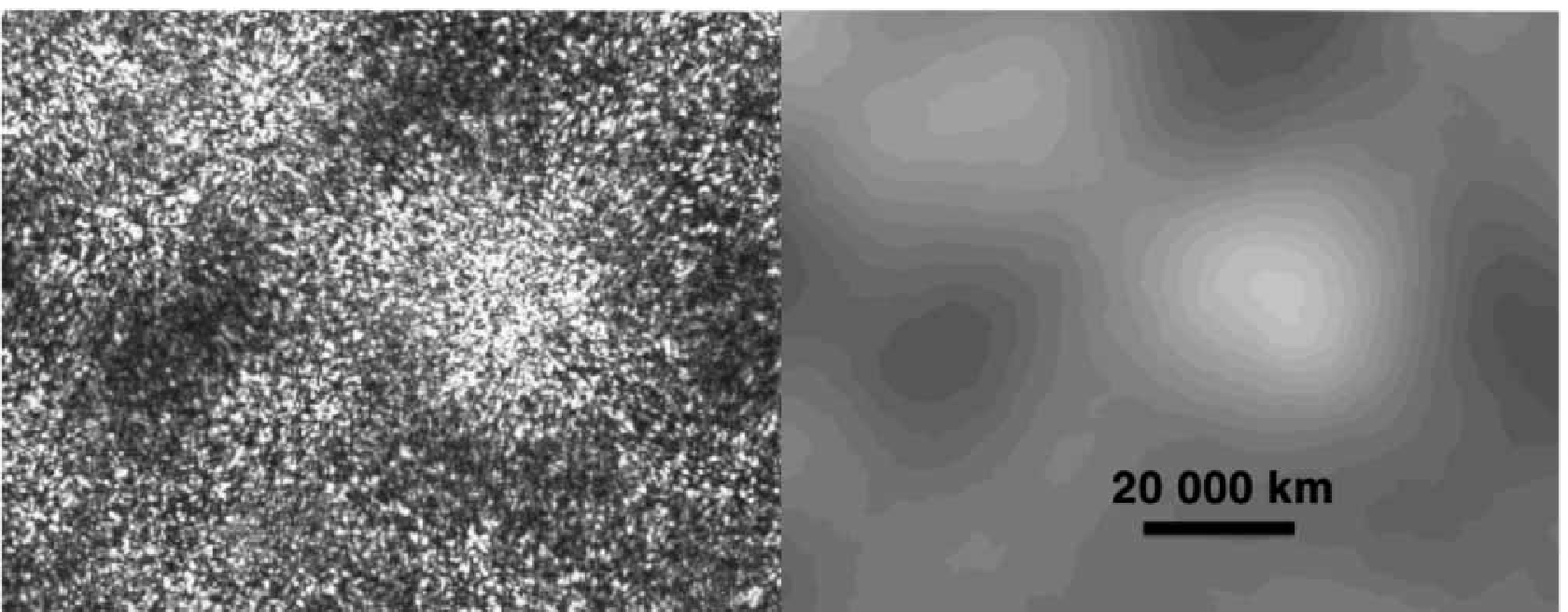}
}
\hspace{0.1cm}
\parbox{3.4cm}{\caption[] 
{\it
Simulated
illumination map at $\lambda=2.16\mu m$ on Earth from a point source (up-left)-
and from a K0V star ($r_s=0.85R_{\odot}$, $M_V=5.9$) at $z_1=8\, kpc$ (right).
The refracting cloud is assumed to be at $z_0=160\, pc$
with a turbulence parameter $R_{diff}(2.16\mu m)=150\, km$.
The circle shows the projection of the stellar disk ($r_s\times z_0/z_1$).
The bottom maps are illuminations in the $K_s$ wide band
($\lambda_{central}=2.162\mu m$, $\Delta\lambda = 0.275\mu m$).
}
\label{simuscint}}
\end{figure}

The illumination pattern from a stellar source of radius $r_s$ is
the convolution of the point-like intensity pattern with the projected
intensity profile of the source
(Fig. \ref{simuscint}, up-right).
%
The cloud, moving with transverse velocity $V_T$
relative to the line of sight, will induce stochastic
intensity fluctuations of the light received from the star
at the characteristic time scale
\begin{equation}
t_{ref}(\lambda) = \frac{R_{ref}(\lambda)}{V_T} \sim
5.2\, minutes\left[\frac{\lambda}{1\mu m}\right]\left[\frac{z_0}{1\, kpc}\right]\left[\frac{R_{diff}(\lambda)}{1000\, km}\right]^{-1}\left[\frac{V_T}{100\, km/s}\right]^{-1}.
\label{dureescint}
\end{equation}
with modulation index $m_{scint.}=\sigma_I/\bar I$ given by
\begin{eqnarray}
m_{scint.} &=& 0.12 \, \left[\frac{\lambda}{1 \mu m}\right] \left[\frac{z_0}{10 pc}\right]^{-1/6} 
                      \left[\frac{R_{diff}(\lambda)}{1000 km}\right]^{-5/6} \left[\frac{r_s/z_1}{R_\odot/10 kpc}\right]^{-7/6}.
\label{xparam} 
\end{eqnarray}
This modulation index decreases when the apparent stellar radius increases.

{\bf Signature of the scintillation signal:}
The first two signatures
point to a propagation effect, which is incompatible with
any type of intrinsic source variability.
\begin{itemize}
\item
Chromaticity:
Since $R_{ref}$ varies with $\lambda^{-1/5}$, one expects
a small variation of the characteristic time scale $t_{ref}(\lambda)$
between the red side of the optical
spectrum and the blue side.
\item
Spatial decorrelation:
We expect a decorrelation between the
light-curves observed at different telescope sites, increasing with
their distance.
\item
Correlation between the stellar radius and the modulation index:
Big stars scintillate less
than small stars through the same gaseous structure.
\item
Location:
The probability for scintillation is correlated with the
foreground gas column-density.
Therefore, extended structures may induce
scintillation of apparently neighboring stars
looking like clusters.
\end{itemize}
{\bf Foreground effects, background to the signal:}
Atmospheric {\it intensity} scintillation is negligible
through a large telescope \cite{dravins}.
Any other atmospheric effect should be easy
to recognize as it affects all stars.
Asterosismology, granularity of the
stellar surface, spots or eruptions
produce variations of very different amplitudes and time scales.
A rare type of recurrent variable stars exhibit emission
variations at the minute scale, but such objects could be identified
from their spectrum.
\section{Preliminary studies with the NTT}
\begin{figure}[h]
\centering
\parbox{7cm}{
\includegraphics[width=7.cm]{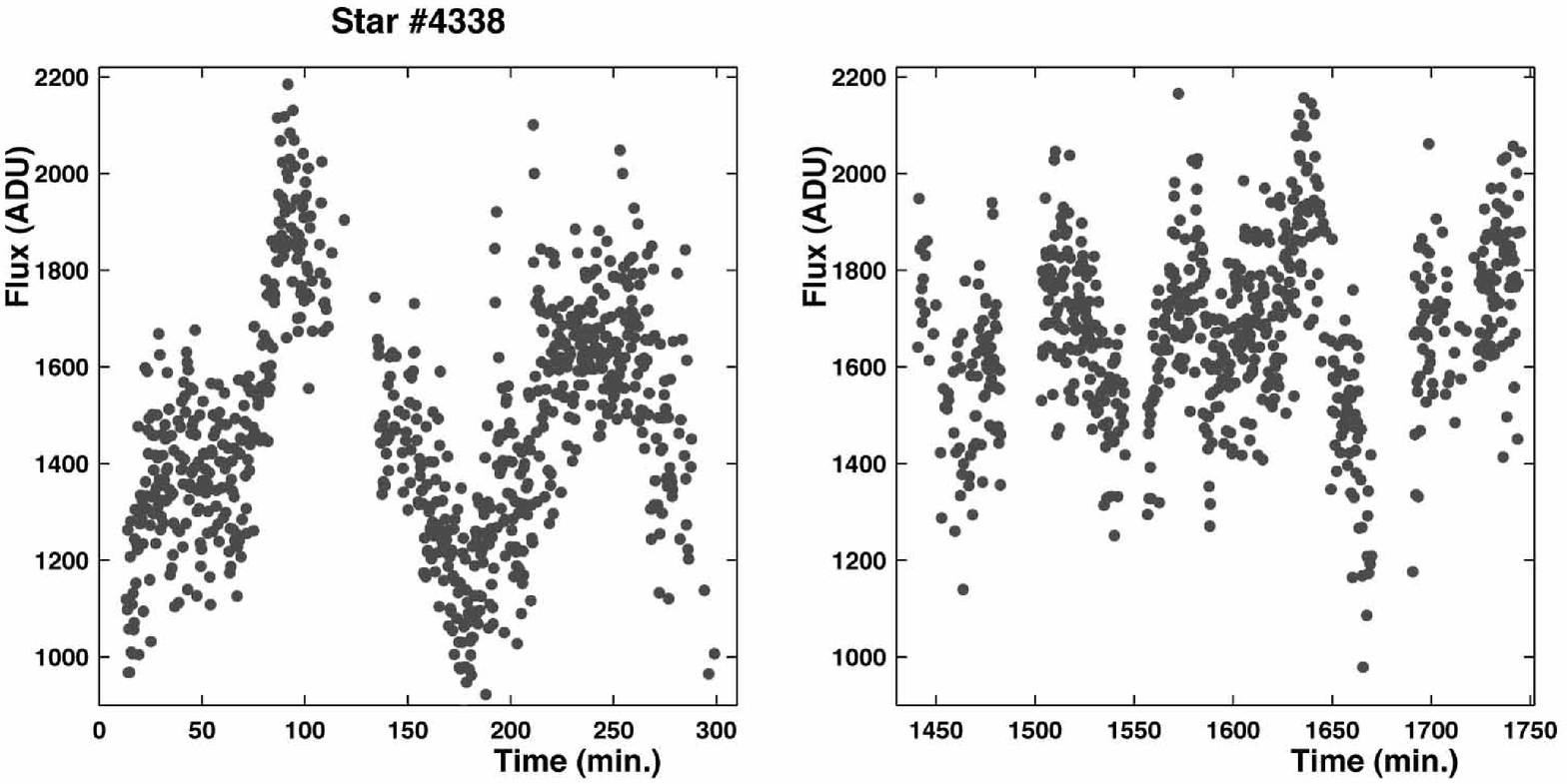}
\caption[]{\it
Light-curves for the two nights of observation (above) and images of the
candidate found toward B68 during a low-luminosity phase (up-right)
and a high-luminosity phase (bottom); North is up, East is left.
\label{candidate}}
}
\hspace{0.3cm}
\parbox{4cm}{
\includegraphics[width=3.5cm]{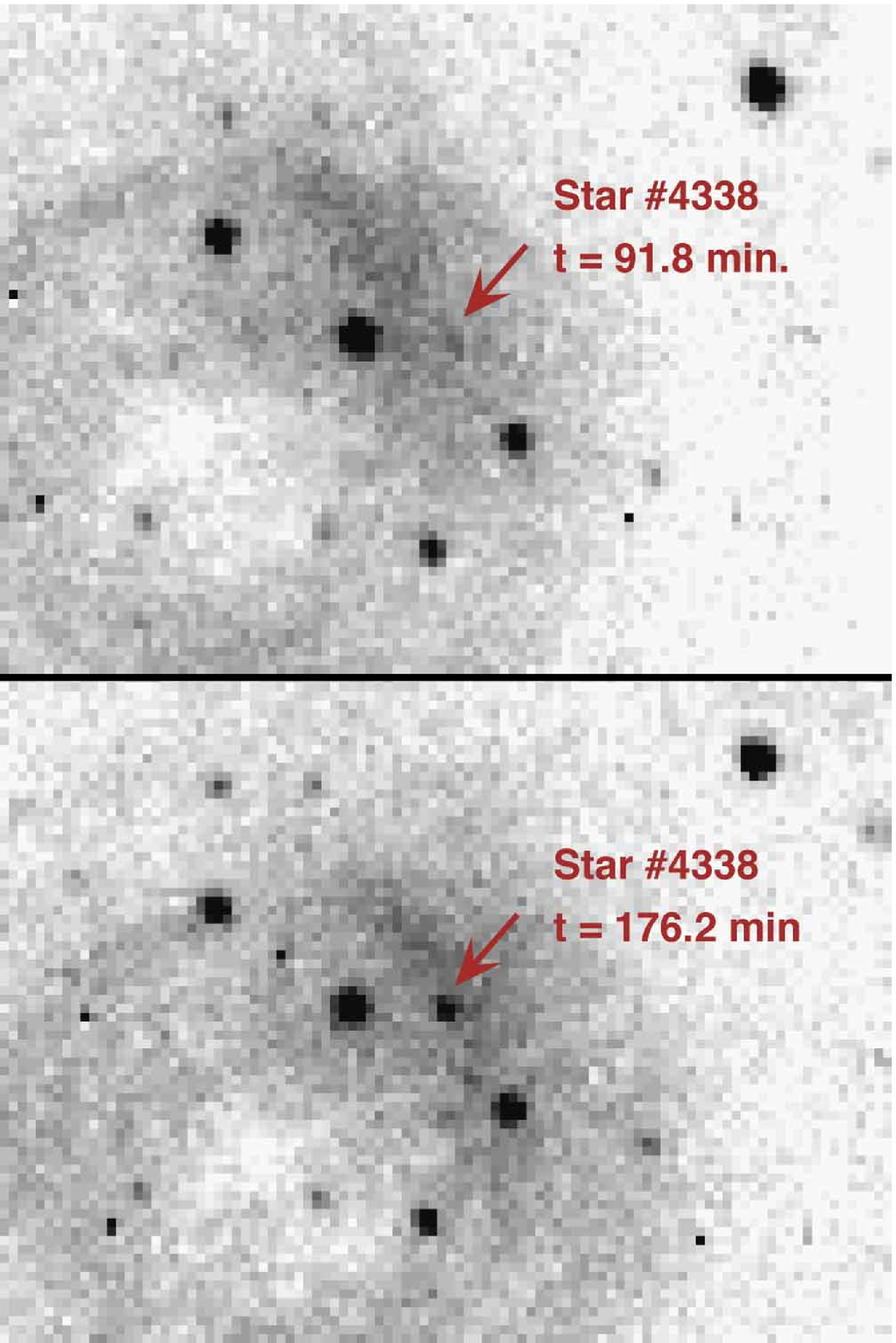}
}
\end{figure}

During two nights of June 2006,
4749 consecutive exposures of ${T_{exp}=10\,s}$
have been taken with the infra-red
SOFI detector in $K_s$ and $J$ through nebulae B68, cb131, Circinus and
towards SMC \cite{resultNTT}.
A candidate has been found towards B68
(Fig. \ref{candidate}), but the poor photometric precision in $K_s$ and
other limitations prevent us from definitive conclusions.
Nevertheless, we can conclude from the rarity of stochastically
fluctuating objects
that there is no significant population of stars that can
mimic scintillation effects, and future searches 
should not be overwhelmed by background of fakes.

\begin{figure}[h]
\centering
\parbox{5.5cm}{
\includegraphics[width=5.5cm,height=3.5cm]{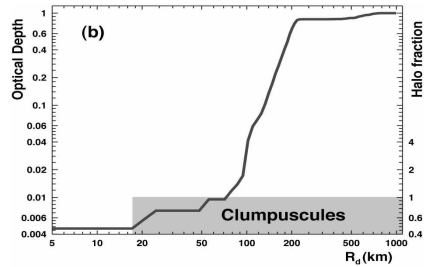}
}
\hspace{0.3cm}
\vspace{-0.3cm}
\parbox{5.cm}{\caption[]{\it
The $95\%\,CL$ maximum optical depth of structures with $R_{diff}(1.25\mu m)<R_d$
toward the SMC. The right scale gives the maximum contribution of
structures with $R_{diff}(1.25\mu m)<R_d$ to the Galactic halo (in fraction);
the gray zone shows the possible region for the hidden gas
clumpuscules expected from the model of \cite{fractal}.
\label{limits}}}
\end{figure}

From the observed SMC light-curves we also
established upper limits
on invisible gaseous structures as a function of their diffusion radius
(Fig. \ref{limits}). This limit, although not really competitive,
already excludes a major contribution of strongly turbulent gas to the
hidden Galactic matter.
These constaints are at the moment limited by the statistics
and by the photometric precision.\\
\vspace{-1cm}
\section{Perspectives}
LSST will be an ideal setup to search for this signature of gas
thanks to the fast readout and to the wide and deep field.
Scintillation signal would provide a new tool to
measure the inhomogeneities and the dynamics of nebulae,
and to probe the molecular hydrogen contribution
to the Milky-Way baryonic hidden matter.

\end{document}